\documentclass[twocolumn,10pt]{article}
\usepackage{authblk}
\usepackage{orcidlink}
\newcommand{\corauthor}[3][]{
    \author[#1,*]{#2}
    \affil[*]{\small Corresponding author: \texttt{#3}}
}

\usepackage[utf8]{inputenc}
\usepackage{hyperref}
\usepackage{graphicx}
\graphicspath{{../figures/}}
\usepackage{setspace}

\usepackage[margin=1.5cm]{geometry}
\usepackage{titlesec}
\titleformat{\subsection}[runin]{\normalfont\bfseries}{}{0.5em}{}[.~]

\usepackage{siunitx}
\usepackage{braket}
\usepackage{amssymb}
\usepackage{amsmath}
\usepackage{sidecap}
\usepackage{changes}
\usepackage{todonotes}
\setuptodonotes{inline, color=orange!30}
\usepackage[version=4]{mhchem}
\usepackage{graphicx,subcaption}
\usepackage[title]{appendix}
\usepackage{makecell}

\DeclareSIUnit{\bohrRadius}{\text{$\text{a}_{\textup{0}}$}}
\DeclareSIUnit\gauss{G}
\DeclareSIUnit\photon{photon}

\newcommand{\Rb}{\ce{^{87}Rb}}

\begin{document}


\title{All-optical measurement of magnetic fields for quantum gas experiments}

\author[1]{Suthep Pomjaksilp\orcidlink{0009-0004-1162-801X}}
\author[1]{Sven Schmidt\orcidlink{0009-0000-5626-2630}}
\author[1]{Aaron Thielmann\orcidlink{0009-0000-0799-9927}}
\author[1]{Thomas Niederprüm\orcidlink{0000-0001-8336-4667}}
\corauthor[1]{Herwig Ott\orcidlink{0000-0002-3155-2719}}{ott@physik.uni-kl.de}
\affil[1]{Department of Physics and Research Center OPTIMAS, Rheinland-Pfälzische Technische Universität Kaiserslautern-Landau, 67663 Kaiserslautern, Germany}

\renewcommand{\abstractname}{\vspace{-6ex}}
\twocolumn[
  \begin{@twocolumnfalse}
    \maketitle
    \begin{abstract}

We present an all-optical method to measure and compensate for residual magnetic fields present in a cloud of ultracold atoms trapped in an optical dipole trap.
Our approach leverages the increased loss from the trapped atomic sample through electromagnetically induced absorption.
Modulating the excitation laser provides coherent sidebands, resulting in $\Lambda$-type pump-probe scheme.
Scanning an additional magnetic offset field leads to pairs of sub-natural linewidth resonances, whose positions encode the magnetic field in all three spatial directions.
Our measurement scheme is readily implemented in a typical quantum gas experiments and has no particular hardware requirements.

    \end{abstract}
    \vspace{3ex}
  \end{@twocolumnfalse}
  ]

\section*{Introduction}

Magnetic field measurement and calibration is a cornerstone in quantum gas experiments.
It allows for precision control of the atomic Zeeman levels, needed in a large variety of applications, like Raman transitions, spinor gases \cite{Stamper-Kurn2013}, spin dynamics and entanglement \cite{Klempt2014}, Feshbach resonances \cite{Chin2010}, or metrology \cite{Schmiedmayer2009,Schmidt2015}.
Traditional methods such as fluxgate magnetometers \cite{Primdahl1979,Du2020}, or high-end methods such as NV centers \cite{Rondin2014} or SQUID sensors \cite{Weinstock1991,Robbes2006} can only measure the field outside the vacuum chamber and microwave equipment, which allows for the precise in-situ measurement of magnetic fields is not always available. 
Atomic magnetometers \cite{Huang2015,Papoyan2016,Azizbekyan2017} use magneto-optical effects to measure the spin precession in a magnetic field.
They are close to  the application discussed here, however, they are often optimized for vapor cell applications. 
In a quantum gas experiment, atomic magnetometry is straightforward to implement since the necessary components, such as lasers, atom detection and magnetic field management are readily available.
It is therefore possible to adapt the core principles of atomic magnetometers to the requirements and constraints of ultracold atom experiments.

Here, we demonstrate an all-optical technique to measure a magnetic field vector in an ultracold atom experiment without prior calibration. 
Our method is easily applicable in most quantum gas and optical tweezer setups.
The method relies on electromagnetically induced absorption (EIA) \cite{Akulshin1998,Lezama1999} on the optical cooling transition in combination with fast chopping of the excitation laser.
The latter creates coherent sidebands at a well defined frequency.
Spectroscopy is performed by scanning an additional magnetic field in all three spatial dimensions.
Thereby, the carrier and the sidebands can be brought into two-photon resonance with neighboring Zeeman sublevels in a $\Lambda$ type setting.
The resonance condition is signaled by an increased absorption due to EIA and corresponding peaks in the spectra.
The measurement signal can be any experimentally accessible observable, which depends on the optical absorption, e.g. the fluorescence or the sample lifetime.
Our method does not require any microwave field and a single laser beam is, in principle, sufficient, to measure the field.

\begin{figure}[t]
    \centering
    \includegraphics[width=\columnwidth]{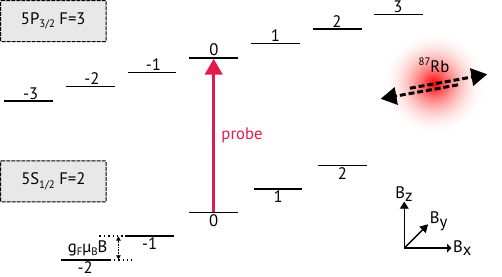}
    \caption{
        We probe the atomic sample with two counter-propagating beams (dashed black arrows) having circular polarization of opposite handedness.
        Three pairs of coils in Helmholtz configuration provide a constant magnetic offset field $\vec{B}=(B_x, B_y, B_z)$. The probing beams are not aligned to any of the coil axis.
        The field $|\vec{B}|$ splits the Zeeman sates of the involved $5\text{S}_{1/2}\text{F}=2$ and $5\text{P}_{3/2}\text{F'}=3$ levels.
        }
    \label{fig:setup}
\end{figure}

In the following, we first introduce the optical tweezer setup, which we use in our experiment.
We then introduce the general measurement scheme, the underlying theory and discuss the obtained spectra.
We close with a discussion of possible applications, extensions and further developments.

\section*{Experimental Setup}
To demonstrate the method, we use mesoscopic atomic ensembles of rubidium atoms trapped in optical tweezers.
In brief, we load \Rb~atoms from the background gas into a 6-axis magneto-optical-trap (MOT), located in a ultra-high vacuum glass cell.
The optical tweezers (waist $w_0\approx\SI{2}{\micro\m}$ at $\SI{1064}{\nano\m}$) are directly loaded from the MOT.
We typically trap 10-20 atoms in a single optical tweezer.
An $\text{NA}=\num{0.4}$ objective, which is used for the creation of the optical tweezers, is also used to collect fluorescence light from the atomic sample both on the D1 and D2 line.
Our setup has an overall photon detection efficiency of approximately \SI{2}{\percent}.
The fluorescence imaging is essentially background free ($<0.01$ photons/second/pixel).
For the spectroscopy, we use a separate set of co-axial counter-propagating beams, which are resonant to the D2 cooling transition $\text{F}=\num{2}\rightarrow \text{F'}=\num{3}$ of \Rb.
Both beams are switched by an acousto-optical modulator (AOM). See Fig.\,\ref{fig:setup} for a sketch of the laser beam geometry and the relevant level scheme.

\begin{figure}
    \centering
    \hfill
    \begin{subfigure}{\columnwidth}
        \caption{}
        \includegraphics[width=\columnwidth]{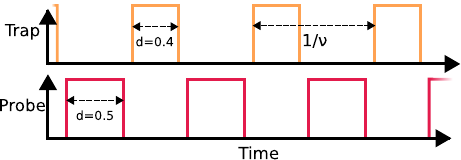}
        \label{fig:chopping_scheme}
    \end{subfigure}
    \hfill
    \begin{subfigure}{\columnwidth}
        \vspace{-.5cm}
        \caption{}
        \includegraphics[width=\columnwidth]{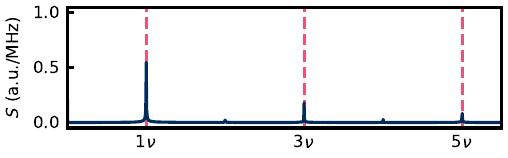}
        \label{fig:beam_spectrum}
    \end{subfigure}
    \caption{
        (a) Alternating sequence of trapping (top) and probing (bottom) the sample.
        Modulating the probe beam with a modulation frequency $\nu$ and a duty cycle $d=0.5$ leads to the formation of AM sidebands. 
        Using a smaller duty cycle for the optical dipole trap $d=\num{0.4}$ omits trap-induced lightshifts during probing.
        (b) Measured frequency spectrum of the probing beam for $\nu=\SI{500}{\kilo\hertz}$ and $d=\num{0.5}$ revealing sidebands with frequency offsets $(2n+1)\nu\;\forall n\in\mathbb{N}_0$. $S$ is normalized to the spectral power of the carrier.
        }
    \label{fig:chopping}
\end{figure}
\begin{figure}
    \centering
    \includegraphics[width=\columnwidth]{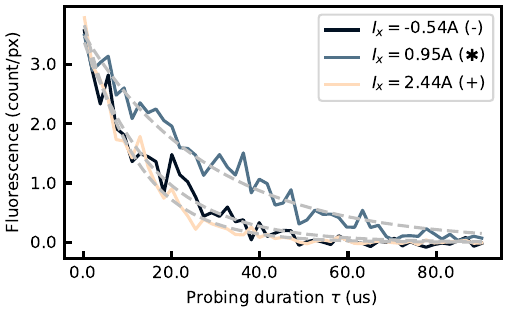}
    \caption{
        Mean D1 fluorescence signal count plotted against the probing duration $\tau$.
        The dashed lines shown an exponential fit to the data.
        The three curves corresponds to the three vertical lines in Fig.\,\ref{fig:initial_loss_ns}.
        }
    \label{fig:decay_ns}
\end{figure}

For the spectroscopy, we chop the probing beams and the optical tweezers with a frequency of $\nu=\SI{500}{\kilo\hertz}$ and a phase shift of $\pi$ with respect to each other (see Fig.\,\ref{fig:chopping_scheme}).
A duty-cycle of $d = \tau_\mathrm{on}\nu = \num{0.5}$ for the probe beams and $\num{0.4}$ for the optical tweezer ensures that the probe beams interact with free atoms without any residual trapping light and resulting lightshifts present.
The modulation of the probing beams leads to the formation of coherent sidebands at frequencies $\pm(2n+1)\nu\;\forall n\in\mathbb{N}_0$ as shown in Fig.\,\ref{fig:beam_spectrum}.
We typically apply \num{250} pulses of the probe beams, before we measure the fluorescence of the remaining atoms on the D1 line.
As the probe beams are resonant to the atomic transition, photon scattering leads to a heating of the atomic sample and a loss of atoms.
We model the time evolution of the atom number $\overline{N}(\tau)$ with an exponential decay
\begin{equation}
    \overline{N}(\tau)=\overline{N_0}\exp(-\gamma\tau)
    \label{eq:decay}
\end{equation}
where $\overline{N}$ denotes the mean number of trapped atoms and $\gamma$ the loss rate.
Fig.\,\ref{fig:decay_ns} shows typical decay curves of the atom number.
The setup is completed by three orthogonal pairs of coils in Helmholtz configuration around the science cell, which can be used to apply external homogeneous magnetic offset fields during the probing sequence.

\section*{Electromagnetically Induced Absorption via Optical Sidebands}
The atomic sample is initially prepared in the $\text{F}=\num{2}$ hyperfine ground state for which the Zeeman levels split according to $\Delta E=m_F g_F \mu_\text{B}|\vec{B}|$ with $g_F=1/2$ in a constant magnetic field $\vec{B}=(B_x, B_y, B_z)$.
For typical ambient fields, this splitting is well below the linewidth of the optical probing transition and direct spectroscopic measurement of the magnetic field is impossible.
However, in three-level systems interferences between excitation pathways can suppress the admixture of the short-lived excited state and thus allow for a spectroscopic resolution below the linewidth of the optical transition.
\begin{figure}
    \centering
    \includegraphics[width=\columnwidth]{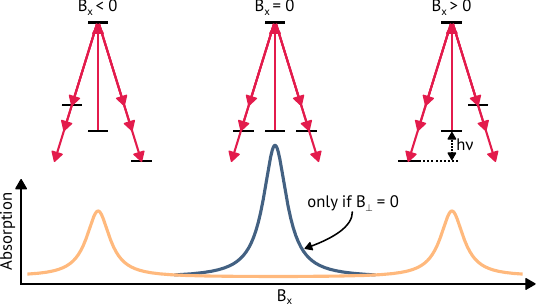}
    \caption{
        Schematics of the $\Lambda$ scheme used in the experiment (top).
        Exemplarily, the carrier of the probe beam drives the $\pi$ transition to the upper hyperfine manifold whereas the sidebands couple to a neighboring $m_F$ state with $\sigma^\pm$ transitions.
        Additional EIA excitation pathways with the same initial and final state are possible but not shown for clarity.
        The bottom half shows the expected absorption spectrum.
        The left peak depicts the situation where the external field $B_x$ brings the $\Delta m_F=\pm1$ level in resonance with the $\pm\nu$ sideband according to Eq.\,\ref{eq:resoncance_criterion}.
        On the right, the $\Delta m_F=\pm1$ level is in resonance with the $\mp\nu$ sideband.
        Both lead to an enhanced absorption via EIA.
        The center peak only occurs if the orthogonal field components $B_\perp$ vanish such that all Zeeman levels are degenerate.
        }
    \label{fig:zeeman_scheme_b_field}
\end{figure}
Here we use EIA in a $\Lambda$ scheme as depicted in Fig.\,\ref{fig:zeeman_scheme_b_field}, where the pump and probe lasers are created by the carrier and the sidebands of the amplitude-modulated excitation laser.
Since the atoms are illuminated by two counter-propagating beams with circular polarization every magnetic transition between the Zeeman levels in the upper $\text{F'=3}$ and the lower $\text{F=2}$ manifold is possible as long as the quantization axis is along one of the three coil axes.
While the two-photon transition in principle allows for ($\sigma^\pm$, $\sigma^\mp$) transitions with $\Delta m_F = \pm 2$ and ($\pi$, $\sigma^\pm$) transitions with $\Delta m_F = \pm1$, a summation over the relevant Clebsch-Gordon coefficients reveals that the latter is up to a factor of eight stronger than the former.
Therefore, it is possible to observe increased photon absorption due to EIA if the two-photon detuning vanishes, $\delta = \frac{g_F \mu_\text{B}}{h}-\nu = 0$, i.e. if the magnetic energy splitting between adjacent Zeeman states matches the modulation frequency creating the sidebands
\begin{equation} \label{eq:resoncance_criterion}
    h\nu = \pm g_F \mu_\mathrm{B} |\vec{B}|
\end{equation}
As depicted in Fig.\,\ref{fig:zeeman_scheme_b_field}, different pathways generated by the sidebands come into resonance depending on the magnetic splitting of the Zeeman levels.
Since EIA also stays effective for small one-photon detunings ($\Delta < \Gamma_\mathrm{5P}$), the magnetic splitting of the excited state Zeeman levels can be neglected and only becomes relevant if a detailed understanding of the observed line strength is necessary.
The increased absorption of photons at $\delta = 0$  directly leads to an increased heating-induced loss rate $\gamma$ from the trap and comprises an easily accessible measurement signal that allows to characterize the external magnetic field.

For the magnetic field, we consider a system of 3 pairs of coils which create a magnetic field with mutually perpendicular components
\begin{align}
    B_n &= \alpha_n I_n
\end{align}
at the position of the atoms.
Here $I_n$ is the current through the respective coil pair and $\alpha_n$ is the system specific conversion factor between current and magnetic field.
We further introduce an unknown offset field $\vec{B}^\mathrm{res} =(B_x^\mathrm{res},B_y^\mathrm{res},B_z^\mathrm{res})$. For convenience, we express the offset field with help of the corresponding current in the coils $B_n^\mathrm{res}=\alpha_n I_n^0$. The modulus of the magnetic field is then given by
\begin{equation} \label{eq:b-field}
    |\vec{B}| = \sqrt{\alpha_x^2(I_x^2 + I_x^0)^2 + \alpha_y^2(I_y^2 + I_y^0)^2 + \alpha_z^2(I_z^2 + I_z^0)^2}
\end{equation}
Combining Eq.\,\ref{eq:resoncance_criterion} and Eq.\,\ref{eq:b-field}, we obtain the following relation for the appearance of the EIA peaks
\begin{align}
    \left(\frac{h\nu}{g_F\mu_\text{B}}\right)^2 &= \alpha_x^2(I_x - I_x^0)^2 + \alpha_y^2(I_y - I_y^0)^2+ \alpha_z^2(I_z - I_z^0)^2
    \label{eq:B_to_nu}
\end{align}

\begin{figure}
    \centering
    \hfill
    \begin{subfigure}{\columnwidth}
        \caption{}
        \includegraphics[width=\textwidth]{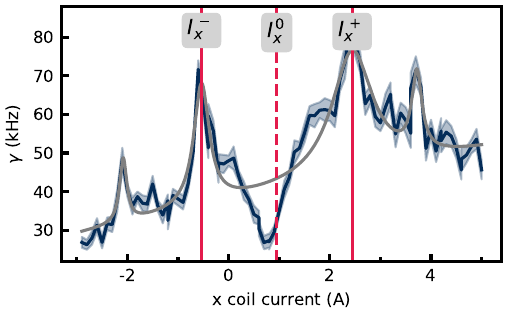}
        \label{fig:initial_loss_ns}
    \end{subfigure}
    \hfill
    \begin{subfigure}{\columnwidth}
        \caption{}
        \includegraphics[width=\textwidth]{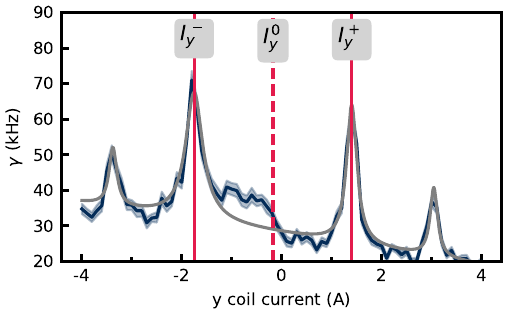}
        \label{fig:initial_loss_ew}
    \end{subfigure}
    \hfill
    \begin{subfigure}{\columnwidth}
        \caption{}
        \includegraphics[width=\textwidth]{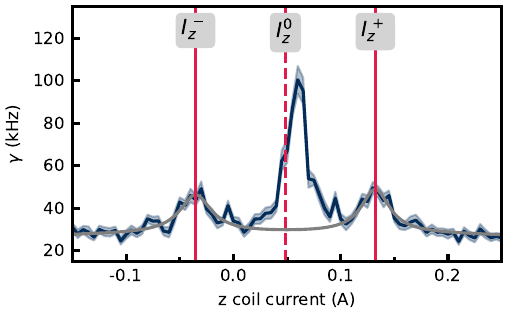}
        \label{fig:initial_loss_ud}
    \end{subfigure}
    \hfill
    \label{fig:all_losses}
    \caption{
        (a) Measured sample loss rate $\gamma$ while scanning the magnetic field in x-direction and keeping the other respective coil currents fixed.
        We determine the positions of the peaks by fitting the spectrum with a multi Lorentzian model (gray line).
        The dotted red line indicates $I_x^0$  whereas the solid red lines mark the peak positions used to determine $I_x$.
        The sample decays for $+$, $-$ and $\ast$ are shown in Fig.\,\ref{fig:decay_ns}. 
        (b), (c) The remaining measurements for the field directions y and z.
        }
\end{figure}

To experimentally probe these field-induced EIA resonances, we create a varying magnetic field in one of the Cartesian directions, e.g. the x-direction, while setting $I_y = I_z = 0$.
For each applied magnetic field, we measure the atom loss from the sample as a function of the excitation pulse duration and extract the decay constant $\gamma$ (Fig.\,\ref{fig:decay_ns}).
Plotting the decay rates as a function of the applied magnetic field reveals the spectrum shown in Fig.\,\ref{fig:initial_loss_ns}.
Except for an overall slope the measured spectrum shows the expected peak structure from Fig.\,\ref{fig:zeeman_scheme_b_field}.
The observed peaks correspond to the $(\pi,\sigma^\pm)$ transitions, while the strongly suppressed $(\sigma^\pm, \sigma^\mp)$ resonances remain hidden in the noise floor of the measurement.
The central peak from Fig.\,\ref{fig:zeeman_scheme_b_field} only appears if also the orthogonal field components $B_y = B_z \approx 0$ approximately vanish. This is only the case if the offset field is small in these two directions or already compensated.
Still, the symmetry of the spectrum allows to identify the current required to compensate the offset field, i.e. $I_x = -I_x^0$ in the x direction, by measuring the center between the two first side peaks.

Compensating the field in the x-direction and conducting an analogue measurement for the y-direction yields the peak structure shown in Fig.\,\ref{fig:initial_loss_ew}.
After compensating the field in the y-direction, the spectrum for the z-direction shown  in Fig.\,\ref{fig:initial_loss_ud} differs in that a central peak  in between the pair of symmetric EIA resonances emerges.
This feature corresponds to a situation where the external fields are zeroed and all Zeeman sublevels are degenerate, as previously observed in \cite{Lipsich2000,Lezama1999a,Akulshin1998}.
In this situation, the carrier alone is able to drive all allowed transitions between $\text{F}=2\rightarrow\text{F'}=3$, greatly increasing the sample absorption and loss rate.
Note that this feature is only visible in the measurement in z-direction since we deliberately compensated the field in the orthogonal directions by setting $I_x = -I_x^0$ and $I_y = -I_y^0$ through the values extracted from the measurements in x- and y-direction.

With the information on the offset currents $I_n^0$, we can apply Eq.\,\ref{eq:B_to_nu}, for the peaks $I_n^\pm$ appearing in the three measurements and obtain a system of three linear equation which we then solve for the conversion coefficients $\alpha_x$, $\alpha_y$ and $\alpha_z$.
Finally, we determine the residual magnetic fields in the experiment as
\begin{equation}
    |B_n^\text{res}| = |\alpha_n I_n^0|
\end{equation}
from the centers of the peak structures.

\begin{table}
    \caption{
        Measured values of the current to field conversion factor $\alpha$ and residual magnetic field $B^\text{res}$ for each coil pair.
        For comparison of the conversion factors, we list the conversion factors $\alpha_\mathrm{sim}$ simulated for our coil geometry.
        To compare the measured residual magnetic field, we list the earth magnetic field $B^\mathrm{earth}$ at the position of our lab according to the world magnetic model \cite{world-magnetic-model}.
        The uncertainties are calculated from the covariance of the fitted peaks. position
        }
    \begin{center}
        \begin{tabular}{ c | r r r r }
        \thead{axis} & \thead{$|\alpha|$\\$(\si{\milli\gauss\per\A})$}  & \thead{$|\alpha_\text{sim}|$\\$(\si{\milli\gauss\per\A})$}  & \thead{$|B^\text{res}|$\\$(\si{\milli\gauss}$)} & \thead{$|B^\text{earth}|$ \\ $(\si{\milli\gauss}$)} \\ 
        \hline
        x & \num{562\pm 17} & \num{680} & \num{498\pm 16} & \num{208} \\  
        y & \num{530\pm 17} & \num{580} & \num{91\pm 3} & \num{6} \\  
        z & \num{10061\pm 17} & \num{9060} & \num{488\pm 5} & \num{482} \\  
        \hline
        \end{tabular}
    \end{center}
    \label{tab:results}
\end{table}
As an  example, we show the conversion factors $\alpha_n$ and residual fields $B_n^\text{res}$ of our setup in Tab.\,Tab.\,\ref{tab:results}.
While the residual field in the z-direction is given by the earths magnetic field, the deviation in the x- and y-direction can be qualitatively explained by a nearby ion-getter pump located \SI{0.5}{\m} beside the science chamber that creates a magnetic field mainly in the horizontal plane.
For comparison of the extracted conversion coefficients, we calculated the conversion factors $\alpha_\text{sim}$ by numerically integrating Biot-Savart's law for our coil geometry.
The measured values show an overall good agreement with the simulated values and deviate by less than \SI{20}{\percent}.
The mismatch can be attributed to the position uncertainties, small remanent magnetic fields of the setup and imperfections in the coil manufacturing.

\section*{Discussion}
This work demonstrates a technique to measure magnetic fields in a quantum gas experiment without prior calibration of the coil system.
The possibility to determine the conversion factors $B_n=\alpha_n I_n$ \textit{on the fly} distinguishes our method from magnetometers based on the Hanle effect while providing a comparable measurement accuracy of $\approx\pm\SI{8}{\milli\gauss}$ \cite{Azizbekyan2017,Papoyan2016}.
While competing methods, which work without prior coil calibration, for example microwave spectroscopy, show a higher accuracy, they typically need additional equipment.
In contrast, our approach needs in essence only resonant laser light and beam chopping. 
Preliminary investigations show that the symmetric structure of EIA peaks in Fig.\,\ref{fig:initial_loss_ns} can even be produced with only one modulated beam.
Our method is also flexible with respect to the detection scheme. 
In principle, any signal that is dependent on the photon scattering rate, such as the sample loss rate, the fluorescence or the transparency of the sample can serve as an observable.
The simplicity and the independence of a prior calibration makes the technique readily available in a typical quantum gas experiments.
Finally, since it is intrinsically Doppler-free in a single beam, our scheme might also find applications in vapor cell magnetometers, using a single probe beam while directly monitoring its transmitted intensity.
The measurement speed is then only limited by the time it takes to perform the magnetic field sweeps.


\section{Data availability} 
The data that support the plots within this paper and other findings of this study are available from the corresponding author upon request.

\section{Code availability} 
The code that support the simulations within this paper and other findings of this study are available from the corresponding author upon request.


\section{Acknowledgements}
We acknowledge financial support by the DFG within the collaborative research center TR185 OSCAR (number 277625399).

\section{Author Contribution}
S.P., S.S. and A.T. performed the experiments. S.P. and S.S. analysed the data. S.P. prepared the manuscript. H.O. and T.N. supervised the experiment. All authors contributed to the data interpretation and manuscript preparation.

\section{Competing financial interests}
The authors declare no competing financial interests.

\end{document}